\documentclass[12pt]{iopart}
\usepackage{graphicx}

\usepackage{color}

\begin{document}
\title[Transient Spin and Phonon Energies from Ultrafast X-ray Diffraction]{Grueneisen-Approach for the Experimental Determination of Transient Spin and Phonon Energies from Ultrafast X-ray Diffraction Data: Gadolinium}
\author{A. Koc$^1$, M. Reinhardt$^1$, A. von Reppert$^2$, M. R\"ossle$^2$, W. Leitenberger$^2$, M. Gleich$^3$, M. Weinelt$^3$, F. Zamponi$^2$ and M. Bargheer $^{1,2}$}

\address{$^1$ Helmholtz-Zentrum-Berlin, Albert-Einstein-Str. 15, 12489
  Berlin, Germany}
\address{$^2$ Institut f\"ur Physik und Astronomie,
  Universit\"at Potsdam, Karl-Liebknecht-Str. 24-25, 14476 Potsdam,
  Germany}
\address{$^3$ Fachbereich Physik, Freie Universit{\"a}t Berlin, Arnimallee 14, 14195 Berlin, Germany}

\ead{bargheer@uni-potsdam.de}

\newcommand{\superscript}[1]{\ensuremath{^{\textrm{#1}}}}
\newcommand{\subscript}[1]{\ensuremath{_{\textrm{#1}}}}

\date{\today}
\begin{abstract}
    We study gadolinium thin films as a model system for ferromagnets with negative thermal expansion. Ultrashort laser pulses heat up the electronic subsystem and we follow the transient strain via ultrafast X-ray diffraction. In terms of a simple Grueneisen approach the strain is decomposed into two contributions proportional to the thermal energy of spin and phonon subsystems.  Our analysis reveals that upon femtosecond laser excitation phonons and spins can be driven out of thermal equilibrium for several nanoseconds.
\end{abstract}

\submitto{\JPCM}
\pacs{75.78.-n, 78.47.J-,78.70.Ck}
\maketitle

\section{Introduction}

Gadolinium has the highest magnetic ordering temperature  among all lanthanides and is ferromagnetic from the Curie temperature $T_C=293$~K down to 4~K, although for bulk Gd a spin-reorientation takes place at $T_{SR}=232$~K. The magnetic moment of the lanthanides mainly arises from the $4f$ shell electrons, and the localized nature of these inner orbitals predestines these elements to test fundamental models of magnetism. Gd is a prototype system for a Heisenberg ferromagnet. Ferromagnetic order is established via the indirect RKKY exchange interaction described by an effective exchange constant $J$.\cite{jens1991a} Gd stands out among the lanthanides for having the largest spin angular momentum of $7.55~\mu_B$ and zero orbital moment. Therefore the interaction with the lattice cannot be explained by crystal field effects.\cite{jens1991a}

\subsection{Thermal expansion coefficient in a Heisenberg ferromagnet}

Above and below the Curie temperature, a large fraction of the thermal energy density $\rho_Q$ in Gd is carried by spin excitations \cite{grif1954}. These spin excitations drive a strong negative thermal expansion below the Curie temperature \cite{darn1963}. An excellent theoretical analysis of the spin-phonon excitation in Heisenberg ferromagnets lead Pytte \cite{pytt1965} to derive the thermal expansion coefficient which is proportional to the specific heat of the phonon- and the spin system, $C_P$ and $C_S$, according to

\begin{equation}
\alpha=\alpha_P+\alpha_S=\frac{1}{9}\frac{1}{d_0^2 K}\left( \frac{3\Gamma_P}{d_0}C_P+\frac{1}{J}\frac{\partial J}{\partial d}C_S\right) .
\end{equation}

$K$ is the elastic modulus and $d_0$ the equilibrium lattice constant. Since the phonon Grueneisen constant $\Gamma_P$ is nearly independent of temperature, the characteristic dip of $\alpha$ around the second order phase transition at $T_C$ indicates that $\frac{\partial J}{\partial d}<0$, i.e. it is the exchange interaction which induces the negative thermal expansion. At low temperatures, the spin-contribution to the specific heat $C_S \sim T^{3/2}$ can be described by magnon excitations and the Grueneisen constant of the spin system can be defined as the logarithmic derivative $\Gamma_S = \frac{\partial \ln J}{\partial \ln V}$ in close analogy to the phonon system $\Gamma_P = \frac{\partial \ln\hbar \omega}{\partial \ln V}$ \cite{lord1967}.
Both Grueneisen constants measure how efficiently the energy density $\rho^Q_{S,P}$ generates stress $\sigma_{S,P}=\Gamma_{S,P} \rho^Q_{S,P}$. Therefore they linearly relate the lattice strain $\epsilon_{S,P} \sim  \sigma_{S,P}  \sim \rho^Q_{S,P}$ to the energy density.

\subsection{Ultrafast magnetization dynamics of Gd thin films}

Spin-excitations and correlations not only govern the  thermo-physical characteristics of Gd. They give likewise rise to the exchange splitting of majority and minority spin bands in the valence band structure and the macroscopic magnetization.\cite{jens1991a} Gd has become one of the most thoroughly studied model systems \cite{Bovensiepen2007} regarding ultrafast optical manipulation of spins and GdFeCo alloys were the first sample system to demonstrate all-optical magnetic switching \cite{Stanciu2007,Radu2011}.

Time-, angle- and energy-resolved photoemission spectroscopy showed that upon optical excitation the exchange splitting decreases within 2~ps \cite{Carley2012}. The response of the majority spin band is delayed by 1~ps and is somewhat slower than the minority spin valence band \cite{Carley2012}. In contrast to the thermal phase transition, the spin polarization of the Gd surface state remains nearly constant within the first picoseconds after laser excitation and decays only slowly within $\tau = 15 \pm 8$~ps \cite{Andres2015}. Photoemission studies of the $4f$ magnetic linear dichroism and the $5d$ exchange splitting showed that their dynamics differ by one order of magnitude, with decay constants of about 14 versus 0.8~ps \cite{Frietsch2015}. The slower picosecond time scale has been attributed to $4f$ spin - lattice coupling \cite{Huebner1996,Melnikov2008,Wietstruk2011}, which determines the spin polarization \cite{Andres2015}.

The notion of thermally driven demagnetization and all-optical switching is prevalent in the recent literature \cite{albi2016,lamb2014}, and already Koopmans et al. suggested that ultrafast demagnetization could be described, disregarding highly excited electronic states, merely considering the thermalized electron system \cite{Koopmans2010}.
The recent experimental work mainly focuses on the ultrafast response of the spins upon optical excitation, although restoring the equilibrium is equally important for the functionality in ultrafast data storage.
Therefore it is important to study how the temperatures of spin and phonon subsystems evolve as a function of time for different starting temperatures. Very recent ultrafast X-ray diffraction experiments on Dysprosium already exploited the strong connection of the lattice expansion and the deposited energy in the spin- and phonon subsystems \cite{repp2016a}. In particular, a persistent non-equilibrium was found with the spin-system remaining hotter than the phonon-system for several nanoseconds, although initially the phonon heating dominated the energy balance.

In this paper we discuss the Grueneisen approach to analyze temperature-dependent time-resolved ultrafast X-ray diffraction data in magnetostrictive systems. As a model system we study a 90-nm thin Gd(0001) film on a W(110) substrate. At low temperatures $T \ll T_C$, our experiments show  a very strong transient contraction of the lattice along the $c$-axis perpendicular to the (0001) surface plane upon laser heating. The contraction is larger than the transient thermal expansion at $T > T_C$, although a considerably larger fraction of the energy deposited in the electron system is transferred to the phonon system.  The phonon system relaxes within 500~ps by heat transport to the substrate. At low temperatures the spin excitations relax on a similar timescale. Close to the Curie temperature, the lattice contraction indicates spin excitations persisting for several nanoseconds. Although the hot electrons had excited both the phonon and the spin system within a few picoseconds, the spin system is largely decoupled from the phonon system, as the equilibration times differ by an order of magnitude near $T_C$. At any initial temperature, the laser excitation drives spin and phonon subsystems out of equilibrium in the sense that they must be described by different temperatures, if the concept of temperature is applicable at all.

\section{Experiment}

The investigated sample is a $D= 90 \pm 10$~nm thick Gd film epitaxially grown at a pressure of $1 \cdot 10^{-10}$~mbar on a (110)-oriented tungsten single crystal substrate.
The tungsten crystal was cleaned following the procedure described in Ref.~\cite{Zakeri2010}. We used homebuilt evaporators and controlled the thickness by a quartz micro balance. First a Gd seed-layer of 10~nm thickness was grown at room temperature and annealed to $400^{\circ}$C.
The completed 90-nm Gd film was annealed to $490^{\circ}$C and showed the low-energy electron diffraction pattern of the hcp(0001) surface. Finally we deposited a polycrystalline Yttrium cap layer of $10 \pm 1$~nm thickness to protect the sample during transport in air to the diffraction experiments. The thickness of the sample was confirmed by acoustic puls echoes measured by ultrafast X-ray diffraction \cite{schi2014c}.

\begin{figure}
  \centering
  \includegraphics[width = 8 cm]{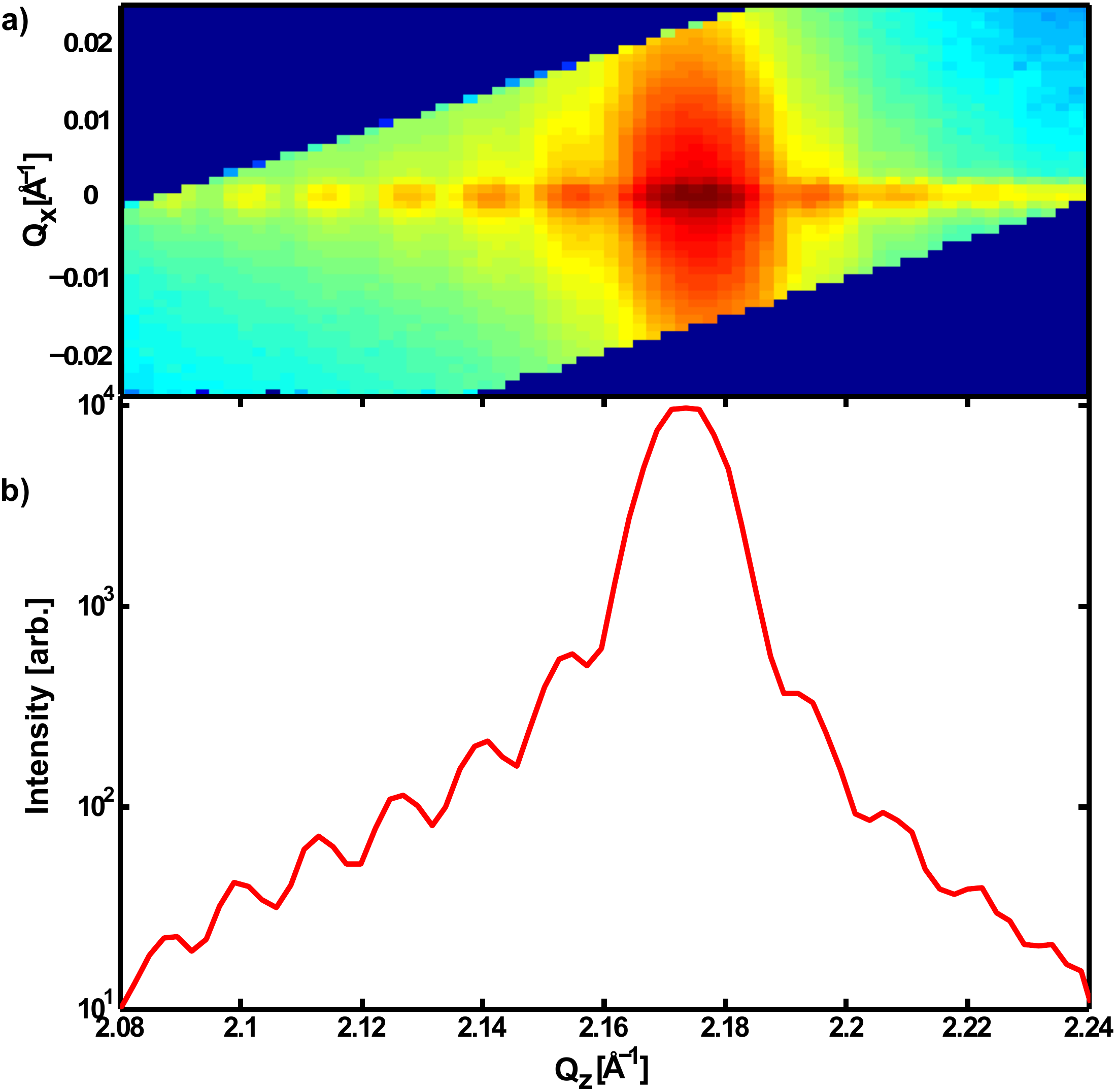}
  \caption{a) Reciprocal space map around the 0002 reflection of Gd. b) Projection of the RSM onto the $q_z$ direction showing pronounced Laue oscillations which indicate a good structural quality.}
\label{fig:RSM}
\end{figure}

Static and time-resolved X-ray diffraction measurements were performed at the XPP experimental station of the storage ring facility BESSY II (Helmholtz-Zentrum-Berlin) \cite{navi2013a,rein2016}.  Figure \ref{fig:RSM}a) shows a reciprocal space map of the Gd film, confirming its good structural quality. In the time-resolved experiments, we determined the lattice strain $\varepsilon (t) = [c(t)-c(t<0)] / c(t<0)$ from projections of the reciprocal space maps (RSM) \cite{schi2013d} of Gd around the (0002) reflection as shown exemplarily in figure~\ref{fig:RSM}b).
The 9~keV hard X-Rays have an attenuation length in Gd of approximately 800~nm and thus probe the entire Gd film. Therefore, the measured Bragg peak shift yields a reliable measure of the average strain in the film. The 250~fs pump pulses have a central wavelength of 1032~nm. The penetration depth of the pump light in Yttrium and Gd is about 23~nm. We emphasize that the Grueneisen concept developed below yields the same average strain, independent of the spatial profile of the deposited energy. The expansion is proportional to the total energy deposited in the spin and the phonon system.
Further details of the setup are given in previous publications \cite{navi2013a,rein2016}.

\section{Results}

\subsection{Spin and phonon Grueneisen coefficient of Gd}

\begin{figure}
  \centering
  \includegraphics [width = 8 cm]{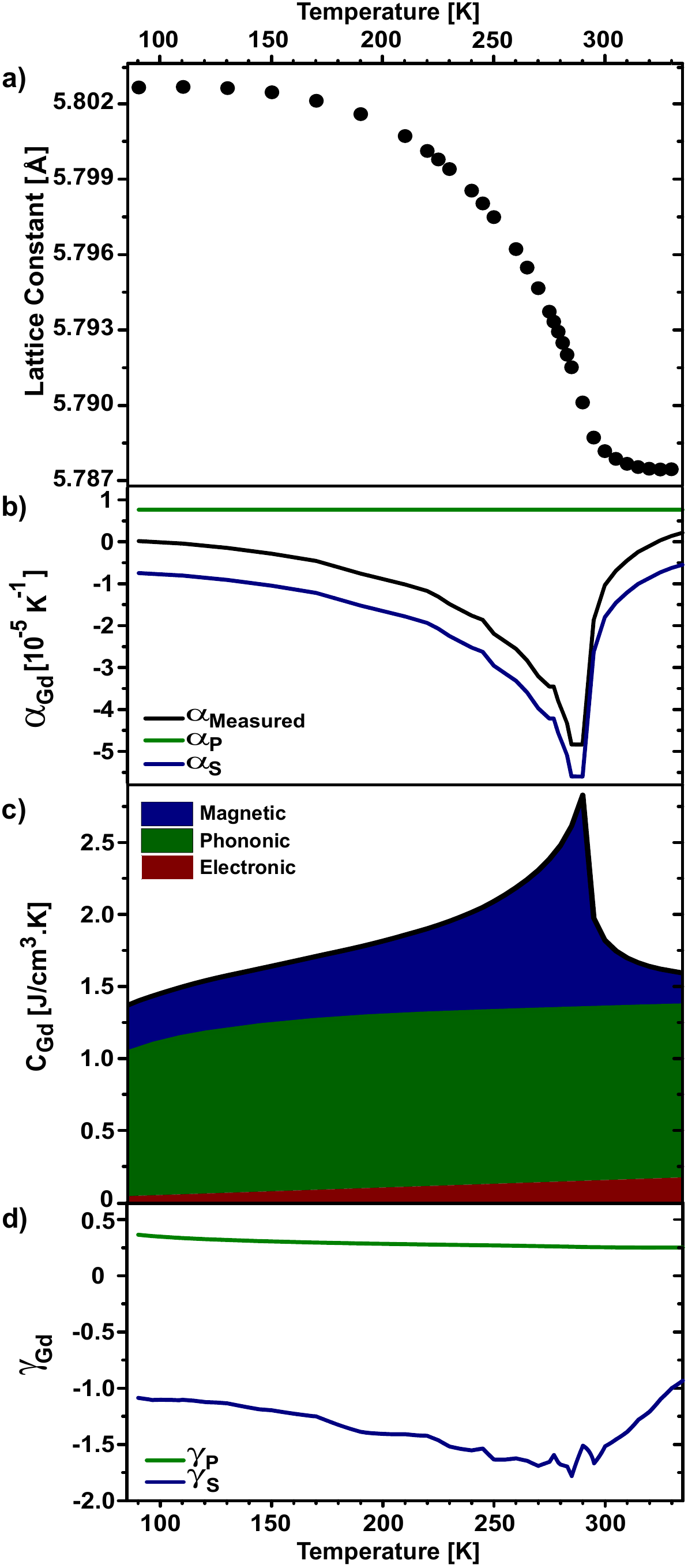}
  \caption {Temperature-dependent paramaters of Gd in equilibrium : a) out-of-plane lattice constant, b) thermal expansion coefficient $\alpha_{Gd}$, c) specific heat capacity $c_{Gd}$ \cite{grif1954} separated into the electronic, phononic and magnetic contributions (see text), d) Grueneisen constant $\Gamma_{Gd}$ derived from the data in panels b) and c).}
\label{fig:static}
\end{figure}

Figure~\ref{fig:static}a) shows the measured temperature-dependent out-of-plane lattice constant $c$ of our Gd(0001) film, which we used to calculate the thermal expansion coefficient $\alpha_{Gd}$ depicted in figure~\ref{fig:static}b).
The strong negative thermal expansion coefficient of Gd is closely related to the specific heat contribution of the ferromagnetically ordered spin-system, as both clearly peak at $293$\,K. For convenience we reproduce the heat capacity $C_{Gd}$ at constant pressure in figure~\ref{fig:static}c) after Ref.~\cite
{grif1954}.
From the relation  $\Gamma = \alpha K/C_{Gd}$
we evaluate the Grueneisen constants $\Gamma_S$ and  $\Gamma_P$ shown in figure~\ref{fig:static}d). The spin and phonon contributions follow from the appropriate decomposition of the thermal expansion coefficient $\alpha_{Gd}=\alpha_{S}+\alpha_{P}$ and the specific heat $C_{Gd}=C_{S}+C_{P}$. In our analysis we neglected the difference between thermal expansion driven by electrons and phonons, since it is only relevant on the few picosecond timescale, when the electron system is significantly hotter than the lattice \cite{repp2016a,nico2011a,nie2006a}. The joint contribution of electrons and phonons to the specific heat was obtained by scaling the phonon contribution to the specific heat values for non-magnetic Lutetium according to the Debye-Temperature \cite{jenn1960a,gers1969}. The phonon contribution of Lu was obtained by subtracting the electron contribution $C_e^{Lu}$ according to the Sommerfeld constant \cite{vent2014}. The corresponding electron contribution $C_e^{Gd}$ for Gd was added to the scaled phonon $C_P^{Gd}$ value to obtain the non-magnetic contributions of Gd shown in figure~\ref{fig:static}c) for electrons (red) and phonons (green). The thermal expansion above 350 K is approximated by the phonon driven expansion in the Debye model and extrapolated to low temperatures. In the relevant temperature range the linear expansion coefficient is essentially constant \cite{darn1963,bars1957}.

\subsection{Time-resolved X-ray diffraction data}

\begin{figure}
  \centering
  \includegraphics[width = 10 cm]{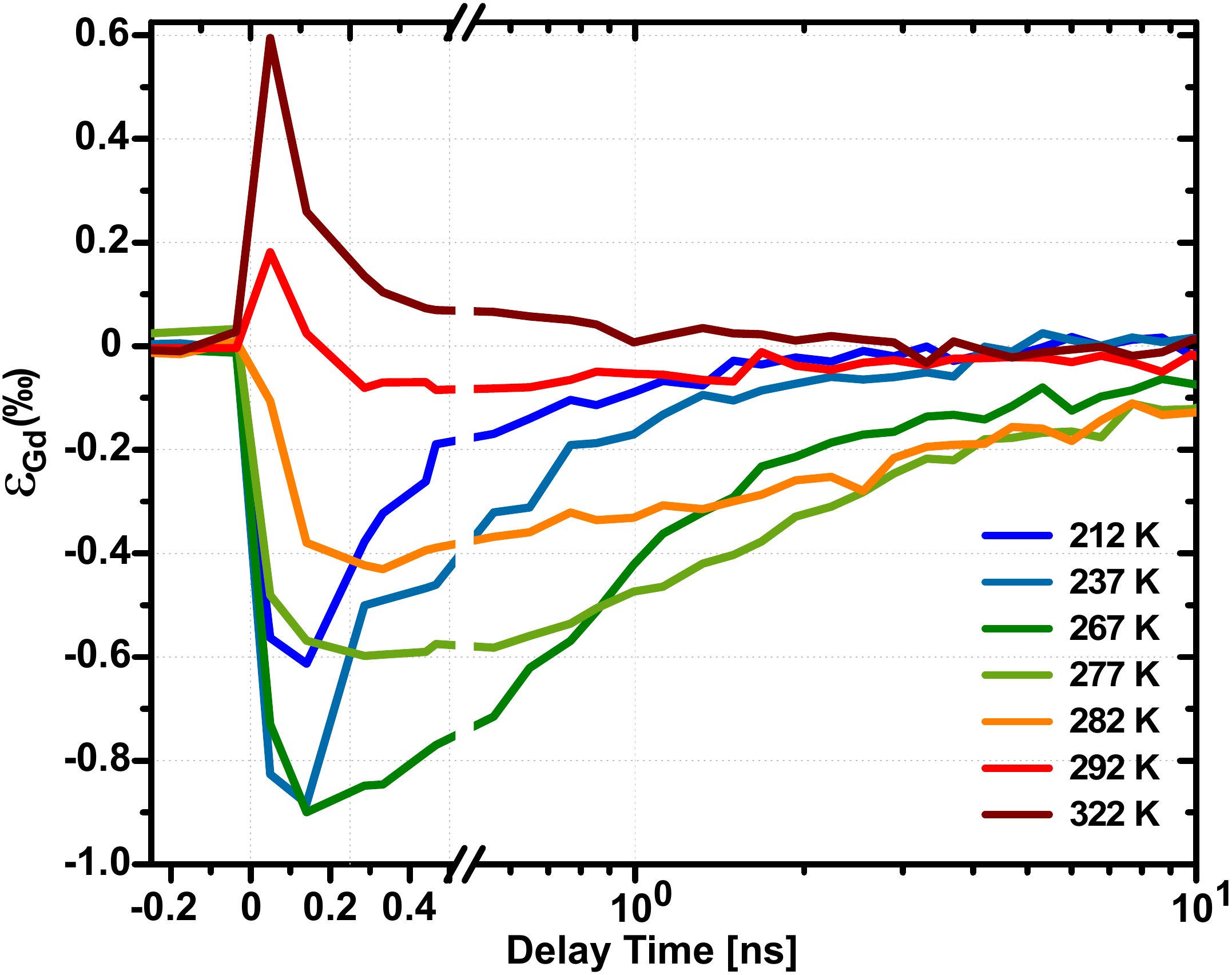}
  \caption{Transient strain $\varepsilon (t)$ in the 90 nm  Gd layer for different initial temperatures $T_{i}$ after ultrafast laser heating at $1.7$\,mJ/cm$^2$ incident fluence. We note that the temperatures given in the legend are the sample temperatures, taking into account the static heat load in the sample imposed by the fluence of $1.7$\,mJ/cm$^2$ at the high repetition rate. The temperature of the sample is increased by 12K, and correspondingly the given $T_i$ values are 12 K larger than the reading of the cryostat's temperature controller.}
\label{fig:UXRDData}
\end{figure}

Figure~\ref{fig:UXRDData} depicts the transient lattice strain in the Gd layer $\varepsilon(t)$ after ultrafast laser heating. For these measurements the initial temperature $T_i$ of the film was tuned across the Curie temperature between 212 and 322\,K, while the excitation was kept constant at an incident fluence of 1.7 mJ/cm$^2$. 

The lattice response at $T_i = 322$~K, i.e. above the Curie temperature, shows an initial expansion of the lattice as expected from phonon heating. Maximum strain is reached within the time-resolution of about 80 ps given by the X-ray pulse duration. The timescale $t_{max}$ for expansion is given by the sound velocity \cite{palm1974} $v_{Gd}=3$\,nm/ps via $t_{max}=D/v_{Gd}=33$\,ps.  The lattice expansion relaxes within about 500 ps via heat diffusion into the W substrate.

All transients recorded for an initial temperature below $T_C$ show a lattice contraction at all time delays. This reflects the intimate coupling between interatomic distances and exchange interaction. The strongest negative expansion of about $-9\cdot 10^{-4}$\,K$^{-1}$ at $T_i = 267$ K even exceeds the positive expansion measured at $T_i = 322$\,K (cf. figure~\ref{fig:UXRDData}).

With $T_i$ approaching $T_C$, the contraction decreases while the relaxation of the transient strain slows down, with considerable contraction persisting longer than 10~ns.  Close to the Curie temperature at $T_i = 292$\,K we observe an initial expansion followed by a contraction for delays $> 150$~ps.

Qualitatively, our data directly prove that a substantial fraction of the energy deposited by the laser heats up the spin system, since only $\rho^Q_S$ drives the negative expansion according to the Grueneisen constant.
The contraction observed for $T_i=292$\,K in figure~\ref{fig:UXRDData} indicates that this is also true close to the Curie temperature, when the magnetic order is almost lost.
Even when the thermal expansion coefficient turns positive above $T_c$ (see figure \ref{fig:static}b), the phonon driven expansion is significantly reduced by spin contributions. By a more detailed analysis below we will show that even the transient at  $T_i=322$\,K is considerably influenced by a contractive stress driven by spin-excitation.

In the following we show that laser-excitation leads to a strong non-equilibrium between spin and lattice subsystems and estimate their different transient temperatures. To this end we can safely assume that within the time resolution of our setup of about 80~ps electron and phonon systems have essentially equilibrated. Therefore we can describe both subsystems by a single temperature and only consider different heating of spins and phonons. From ellipsometry measurements of the Y capping layer we know that the energy deposited by the laser pulse is independent of the initial temperature in the relevant temperature range.

First we attempt to analyze the data in an equilibrium model by assuming that spins and phonons have the same temperature at all times. This enforces the static lattice constant $c(T)$ depicted in figure~\ref{fig:static}a) to map the transient temperature $T(t)$ via $c(t)=c(T)$. Likewise the transient strain would mimic the transient temperature via $\epsilon(t) = \epsilon(T)$.

\begin{figure}
	\centering
	\includegraphics[width = 14 cm]{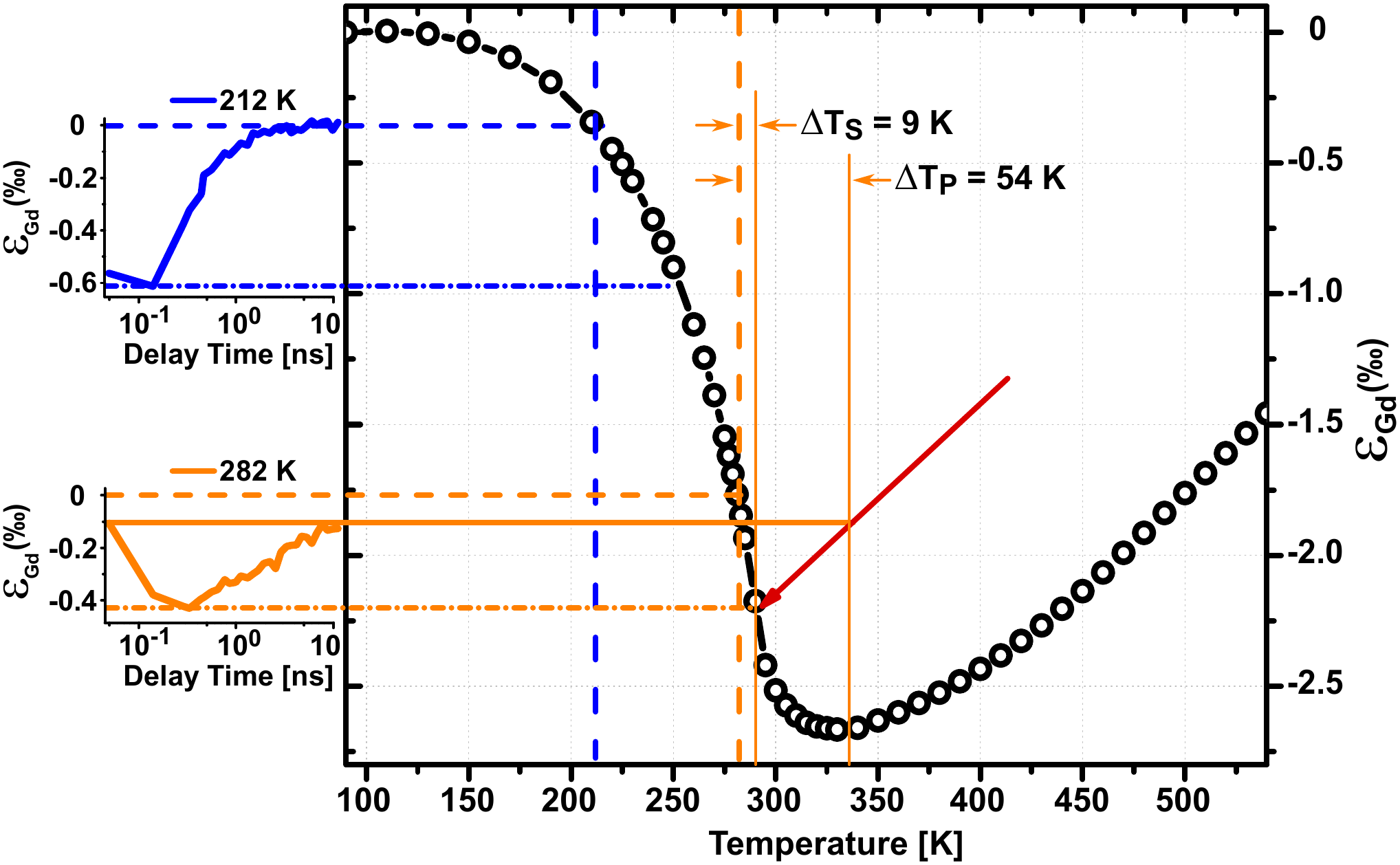}
	\caption{Illustration of the temperature rise in the Gd film after ultrafast laser excitation. The insets show the transient lattice response $\varepsilon (t)$ for two selected initial temperatures $T_i=212$\,K (blue) and 282\,K (orange) reproduced from figure~\ref{fig:UXRDData}. The solid circles represent the measured lattice strain from static heating experiments (figure~\ref{fig:static}a) extrapolated by using high temperature literature values \cite{darn1963}. The thick dashed lines indicate the initial temperature $T_i$ and the corresponding $\varepsilon = 0$. The dash-dotted lines show the maximum contraction, which yields a lower estimate of the temperature rise $\Delta T_S$ of the spin excitations. The solid lines highlight the initial non-equilibrium response that explains the tiny initial contraction at $t=50$\,ps resulting from a near cancellation of the contractive stress from heating spins by $\Delta T_S=9$K and the expansive stress from heating the phonon system by $\Delta T_P=54$\,K. The red arrow symbolizes the exclusive cooling of the phonon system towards the equilibrium at $\Delta T_S=\Delta T_P=9$\,K. The real dynamics will deviate from the red arrow by simultaneous additional $\Delta T_P=3\Delta T_P$ which leads to the same strain according to the Grueneisen coefficients.}
	\label{fig:GA}
\end{figure}

We show for three different examples that this equilibrium approach is incorrect:

(i) Figure~\ref{fig:GA} illustrates two situations for different starting temperatures $T_i$.  If spins and phonons were in equilibrium, the maximum transient strain of $-4.3\cdot 10^{-4}$ observed for $T_i = 282$\,K would correspond to a temperature rise of $9$\,K.  In contrast, the maximum strain for $T_i=212$\,K is $-6.1\cdot 10^{-4}$, which would reflect a temperature rise of $32$\,K. These values can be directly read from the dashed and dash-dotted orange and blue lines in figure~\ref{fig:GA}. The maximum temperature change extracted from the change in $\epsilon(T)$ varies by about a factor of 4 for the two different starting temperatures, although the optical absorption coefficient is constant and consequently the same amount of energy is deposited.
We can safely conclude from the observation at $T_i = 212$\,K that the energy deposited by the laser pulse must at least lead to a temperature rise of $\Delta T > 32$\,K. At $T_i = 282$\,K the total heat capacity of Gd is at most $25$~\% larger than at $T_i = 212$\,K. Therefore, the temperature jump would at least be 24 K, contradicting our simple equilibrium description.

\begin{figure}
  \centering
  \includegraphics[width = 10 cm]{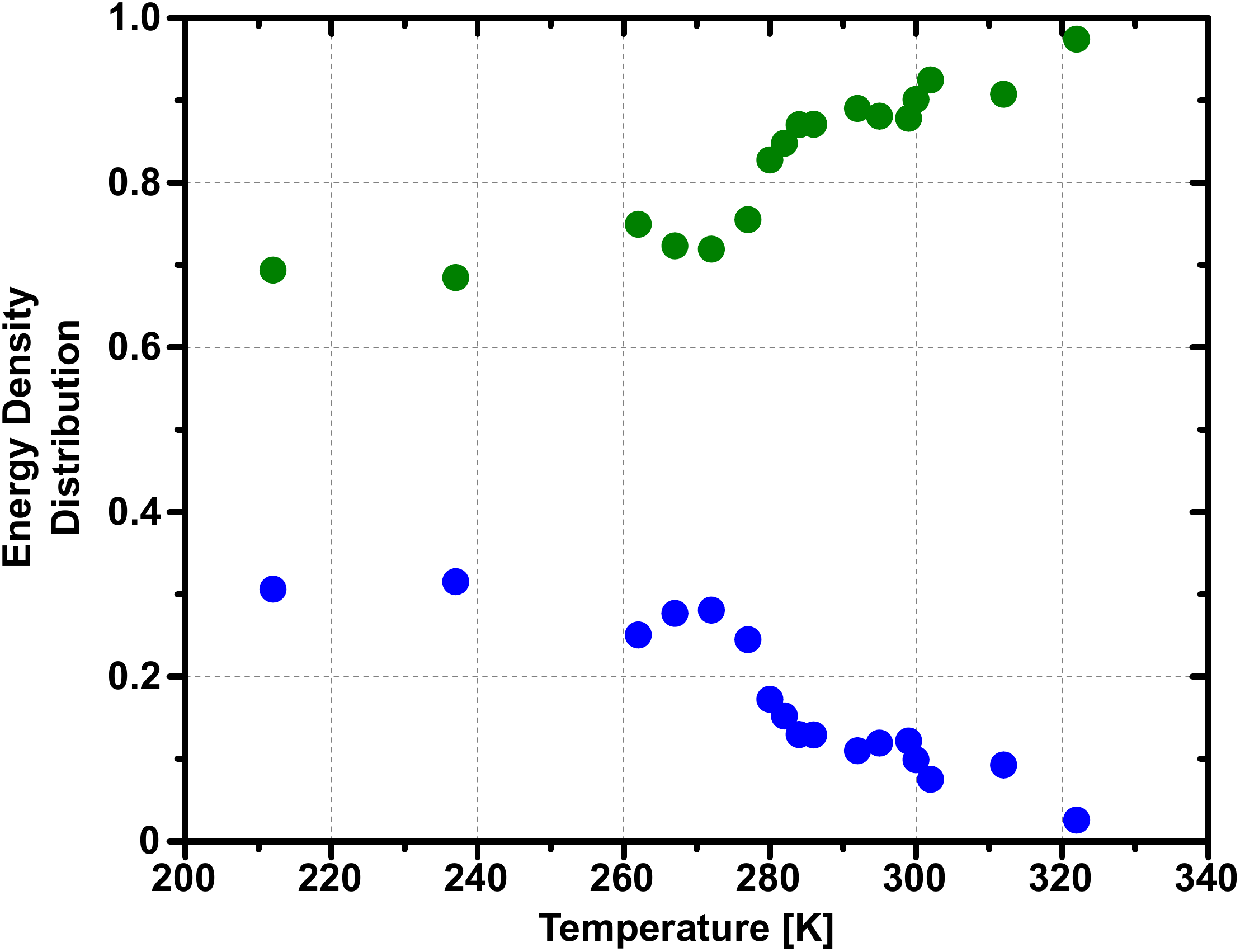}
  \caption{Fraction of energy deposited in the spin system (blue) and in the electron-phonon system (green) as a function of initial temperature $T_i$. Note that at $T>T_C$ about $10\%$ of the energy leads to spin excitations.}
\label{fig:SvsP}
\end{figure}

(ii) Moreover, if we consider the situation at a delay of $50$\,ps (solid orange line in figure~\ref{fig:GA}), the failure to assume $T_S = T_P$ becomes even more obvious. The lattice response shows a strain of $-1\cdot 10^{-5}$ corresponding to a temperature rise of only 2~K. In fact this can only be explained in a non-equilibrium model by assuming a simultaneous heating of spin- and phonon systems with an almost exact cancellation $\alpha_S \Delta T_S \approx \alpha_P \Delta T_P$ of the lattice contraction and expansion.  
If we write this relation in terms of the energy densities it follows $-\Gamma_S \rho^Q_S = \Gamma_P \rho^Q_P = \Gamma_P (\rho^Q_{Gd}-\rho^Q_S)$. From the latter identity we can immediately conclude that the spin system initially takes the fraction $\rho^Q_S/\rho^Q_{Gd} = \Gamma_P / (\Gamma_P-\Gamma_S) = 14$~\% of the total deposited energy $\rho^Q_{Gd}$ at $T_i = 282$\,K. This analysis is robust, because strain and energy density are linearly related, and it implies a similarly large difference in the temperature rise of the spin and phonon system, since the heat capacity $C_S$ at 282\,K is only 20$\%$ smaller than $C_P+C_e$. Figure~\ref{fig:SvsP} shows the distribution of energy density $\rho^Q_S$ and $\rho^Q_P$ for different initial temperatures, determined 50\,ps after the excitation.

(iii) We get the third important estimate from the temperature rise required to rationalize the contraction of $-4.3\cdot 10^{-4}$ for $T_i = 212$\,K. In equilibrium we have to assume at least a temperature rise  $\Delta T > 32$\,K to account for the contraction. In non-equilibrium we can, however, also explain the observation by a larger temperature rise $\Delta T_S$ of the spin system, if we assume that the phonon system receives even more energy. We explain this situation graphically in figure~\ref{fig:GA} for the example of $T_i = 282$\,K. At $t=50$\,ps, the dynamics start at a small negative strain level indicated by the solid orange line. This is the strain level given by an approximate balance of expansive and contractive stresses.

The dash-dotted orange line indicates the maximum negative strain that the transient attains at about 300 ps. At this time spin and phonon systems must at least still be heated by $9$\,K. In order to explain the strain balance just after excitation, we have to assume an excessive heating of the phonon system beyond the temperature of the spin system. The red arrow indicates the slope at which the phonon heating leads to strain according to the high-temperature thermal expansion. If we assume an initial temperature jump of the spin system by $\Delta T_S=9$\,K then the phonon systems must exhibit an initial temperature jump of $\Delta T_P=54$\,K to compensate the negative strain of the spin system. Therefore, the red arrow would indicate the path of the system towards equilibrium. Initially, it corresponds to exclusive phonon cooling (along the red arrow) until $\Delta T_P=\Delta T_S=9$\,K, followed by cooling along the static equilibrium.

These three examples show that we can prove the non-equilibrium of spin and phonon subsystems directly from the UXRD transients without detailed calculations. In the following we will discuss an analytic decomposition of the transient signal for any $T_i$, which for example shows that at $T_i=282$\,K, the phonons are initially heated by $\Delta T_P=70$\,K, whereas the spins heat only up by $\Delta T_S=15$\,K.

\section{Discussion}

\subsection{Data analysis in the two-thermal-energies-model (TTEM)}

For an analytic decomposition of the observed lattice strain, we have derived the macroscopic Grueneisen constants $\Gamma_{P,S}=\frac{\alpha_{P,S}(T)\,K_{Gd}}{C_{P,S}(T)}$  in figure~\ref{fig:static}d). $K_{Gd}$ is the effective elastic constant of Gd, which depends only weakly on the temperature. The laser-induced change of the energy densities $\rho^Q_{P,S}$ generates the stresses $\sigma_{P,S} = \Gamma_{P,S} \rho^Q_{P,S}$ which superimpose to yield the proportional strain:

\begin{equation}
\varepsilon_{Gd}=\frac{\sigma_{P}+\sigma_{S}}{K_{Gd}}=\frac{1}{K_{Gd}}(\Gamma_{P} \rho^Q_{P}+\Gamma_{S} \rho^Q_{S})
\label{eq:stressstrain}
\end{equation}

We have measured $\varepsilon_{Gd}(t)$ and know the parameters $\Gamma_{S,P}$ and $K_{Gd}$ \cite{palm1974}. In order to find the values of $\rho^Q_{S}$ and $\rho^Q_{S}$ we use $\rho^Q_{Gd}(t)=\rho^Q_{P}(t)+\rho^Q_{S}(t)$. We get $\rho^Q_{Gd}(0ps)$ from the absorbed laser fluence, which we have calibrated carefully by measuring the transient thermal expansion of materials which have only conventional heat expansion. As an approximation for $t>0$\,ps, we assume that the heat transport in Gd is dominated by the electrons and approximately independent of the temperature \cite{nell1969}. Since for $T_i=322$\,K the energy is essentially fed into the phonon system, we take the observed dynamics also as an estimate for the phonon cooling at lower $T_i$.

\begin{figure}
  \centering
  \includegraphics[width = 10 cm]{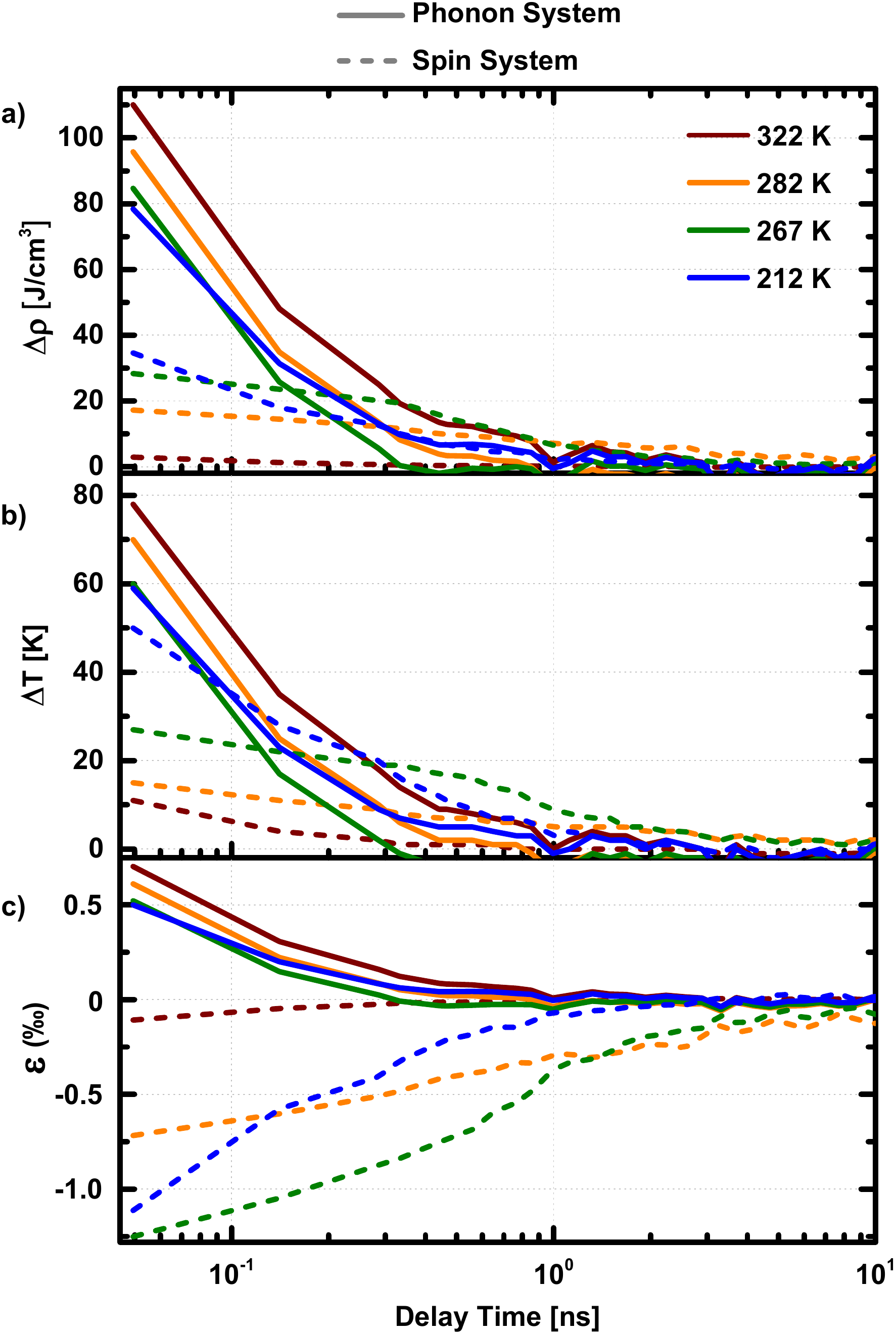}
  \caption{Result of the data analysis for selected initial temperatures. Solid and dashed lines represent the results for phonon and spin systems, respectively. a) Transient energy density, b) temperature change and c) transient strain.}
\label{fig:DelT}
\end{figure}

Figure~\ref{fig:DelT} summarizes the results of the analytic decomposition. Figure~\ref{fig:DelT}a) shows the energy densities in the spin and phonon systems. For $T_i=267$\,K as much as $30\%$ of the deposited energy enters the spin system. The phonon system cools faster than the spins, and already after 200\,ps less than $50\%$ of the energy resides in the phonons. For $T_i=322$\,K the contribution of the spin system starts with less than 10$\%$.
The transport of heat and the interconversion of energy between subsystems should be governed by the temperature, which is plotted in figure~\ref{fig:DelT}b). The spin-temperature increase is reduced at temperatures just below $T_c$. This in part explains why the cooling of the spins for $T_i=267$\,K and 282\,K is slowed down. For higher initial temperature, the fraction of energy initially deposited in the spin system is reduced (cf. figure~\ref{fig:SvsP}), and the specific heat increases near the phase transition. Both facts reduce the change of the spin temperature $\Delta T_S$ and hence it takes a longer time until the phonon temperature $T_P$ has cooled below $T_S$. 

\subsection{Comparison of the results to recent literature}
 In the results for all initial temperatures $T_i<T_C$ below the phase transition, we observe the phonon system is initially heated more than the spin system. Not only is the contribution to the energy density higher ($\rho^Q_P>\rho^Q_S$, cf. figure~\ref{fig:DelT}a), but also the temperature rise $\Delta T_P>\Delta T_S$ (cf. figure~\ref{fig:DelT}b). The phonon system cools faster than the spin system, i.e. after a time $t_{eq}$, they reach equal temperatures ($\Delta T_P=\Delta T_S$), after several hundreds of picoseconds. However, the phonon system keeps cooling faster and therefore the temperature rise is inverted ($\Delta T_P<\Delta T_S$). This inversion of the temperatures observed for all transients below $T_C$ shows that the spin system is decoupled from the electron-phonon system on the 100\,ps - 1\,ns timescale, although the initial transfer of energy from electrons to phonons and spins is very rapid. Note that $t_{eq}$ is smaller for low temperatures and the inversion is more pronounced towards the phase transition. Probably fluctuations at the phase transition reduce the efficient loss of spin energy, because this at the same time requires a decreasing spin-entropy, when the spins reorder. The persistence of the spin-excitations is much more pronounced in Dysprosium, where the spin reordering has been observed to take several tens of nanoseconds \cite{repp2016a,koc2017}. We believe that the ordered spin system around the laser-heated spot dictates a direction along which the spins reorder in the ferromagnetic Gadolinium, whereas the cooling in antiferromagnetic Dysprosium lacks a preferential direction for establishing order.
 For a precise comparison to the ultrafast demagnetization experiments of Gd \cite{Carley2012,Frietsch2015,Sultan2011}, additional measurements at the same sample temperature around $100$~K would be helpful. However, in the UXRD experiments at the lowest temperature of 212~K measured in our study, the phonon temperature increase decays to half of its value within 120 ps and the spin temperature rise requires about 200 ps to decay to its half value. Taking into account the larger thermal conductivity at 100 K, the results are consistent with relaxation times of about 80 ps reported in the literature \cite{Carley2012,Frietsch2015,Sultan2011}.

We emphasize that qualitatively our main conclusion can be directly drawn from the data for $T_i =292$~K in figure~\ref{fig:UXRDData}. We start just below the phase transition, and initially the phonon driven expansion dominates the signal, whereas after about 150 ps the persistent spin excitations yield the contraction. If both systems would dissipate energy at the same rate, such a sign change of the strain should not be observed. These results clearly call for a microscopic simulation of the heat transport, which takes into account the interconversion of heat between electron-, phonon-, and spin-degrees of freedom. From a macroscopic point of view it is somewhat puzzling, that the heat transport is dominated by the electron system, although the electron system contributes much less to the specific heat than phonons and spins. Only the electronic heat transport cooling the spin system seems to be reduced near the phase transition. This could be related to changes of the band structure near the Fermi surface \cite{doeb2010} and to the loss of RKKY interaction due to the spin disorder.

\section{Conclusion}

We derived the nearly temperature-independent Grueneisen coefficients of the spin- and phonon systems in the temperature range between $212$\,K and $322$\,K, $\Gamma_{S}\approx -1.54$ and $\Gamma_P=0.26$ from literature values of the strongly temperature dependent specific heat $C_{Gd}$ and the thermal expansion coefficient $\alpha_{Gd}$ measured in the thin film.

By analytically separating the phonon- and spin-contributions ($\rho^Q_{S,P}$) to the measured transient signal, we find that the fraction of energy deposited in the spin system decreases from $33\%$ at 212\,K to about $9\%$ around $T_c$. The analysis is robust, because the measured strain $\varepsilon$ depends linearly on both energy densities $\rho^Q_{S}$ and $\rho^Q_{P}$. Even above $T_c$, in the absence of long-range ferromagnetic order, there is a considerable negative contribution of the spin excitations to the static and transient strain. Although the optical excitation of the valence electrons rapidly couples the energy into the phonon and spin system within few ps, the cooling and reordering of the spin system is much slower than the cooling of the phonon system.

We expect that this study triggers additional work on the experimental and theoretical level, in order to obtain a full understanding of the coupling mechanisms and the heat transport pathways in non-equilibrium situations, which are regularly met in ultrafast magnetic switching experiments.

\section{Acknowledgement}

We are grateful to financial support by the BMBF and the Helmholtz Virtual Institute \textit{Dynamic Pathways in Multidimensional Landscapes}.

\section*{References}
\bibliographystyle{jphysc}

\begin{thebibliography}{10}

\bibitem{jens1991a}
Jensen J and Mackintosh A.~R, {\em {Rare Earth Magnetism - Structures and
  Excitations}} (Clarendon Press, ADDRESS, 1991).

\bibitem{grif1954}
Griffel M, Skochdopole R.~E, and Spedding F.~H, {\em Physical Review} {\bf 93},
   657  (1954).

\bibitem{darn1963}
Darnell F, {\em Physical Review} {\bf 130},  1825  (1963).

\bibitem{pytt1965}
Pytte E, {\em Annals of Physics} {\bf 32},  377   (1965).

\bibitem{lord1967}
Lord A, {\em Journal of Physics and Chemistry of Solids} {\bf 28},  517
  (1967).

\bibitem{Bovensiepen2007}
Bovensiepen U, {\em J. Phys.: Cond. Matt.} {\bf 19},  083201  (2007).

\bibitem{Stanciu2007}
Stanciu C {\it et~al.}, {\em Phys. Rev. Lett.} {\bf 98},  207401  (2007).

\bibitem{Radu2011}
Radu I {\it et~al.}, {\em Nature} {\bf 472},  205  (2011).

\bibitem{Carley2012}
Carley R, D\"obrich K, Frietsch B, Gahl C, Teichmann M, Schwarzkopf O, Wernet
  P, and Weinelt M, {\em Phys. Rev. Lett.} {\bf 109},  0574012  (2012).

\bibitem{Andres2015}
Andres B, Christ M, Gahl C, Wietstruk M, Weinelt M, and Kirschner J, {\em Phys.
  Rev. Lett.} {\bf 115},  207404  (2015).

\bibitem{Frietsch2015}
Frietsch B, Bowlan J, Carley R, Teichmann M, Wienholdt S, Hinzke D, Nowak U,
  Carva K, Oppeneer P, and Weinelt M, {\em Nature Communications} {\bf 6},
  8262  (2015).

\bibitem{Huebner1996}
H\"ubner W and Bennemann K.~H, {\em Phys. Rev. B} {\bf 53},  3422  (1996).

\bibitem{Melnikov2008}
Melnikov A, Prima-Garcia H, Lisowski M, Gie{\ss}el T, Weber R, Schmidt R, Gahl
  C, Bulgakova N.~M, Bovensiepen U, and Weinelt M, {\em Phys. Rev. Lett.} {\bf
  100},  107202  (2008).

\bibitem{Wietstruk2011}
Wietstruk M, Melnikov A, Stamm C, Kachel T, Pontius N, Sultan M, Gahl C,
  Weinelt M, D{\"u}rr H.~A, and Bovensiepen U, {\em Phys. Rev. Lett.} {\bf
  106},  127401  (2011).

\bibitem{albi2016}
Albisetti E {\it et~al.}, {\em Nature nanotechnology}  (2016).

\bibitem{lamb2014}
Lambert C.-H {\it et~al.}, {\em Science} {\bf 345},  1337  (2014).

\bibitem{Koopmans2010}
Koopmans B, Malinowski G, Dalla~Longa F, Steiauf D, F{\"a}hnle M, Roth T,
  Cinchetti C, and Aeschlimann M, {\em Nature Materials} {\bf 9},  259  (2010).

\bibitem{repp2016a}
von Reppert A, Pudell J, Koc A, Reinhardt M, Leitenberger W, Dumesnil K,
  Zamponi F, and Bargheer M, {\em Structural Dynamics} {\bf 3},    (2016).

\bibitem{Zakeri2010}
Zakeri K, Peixoto T, Zhang Y, Prokop J, and Kirschner J, {\em Surf. Sci.} {\bf
  604},  L1  (2011).

\bibitem{schi2014c}
Schick D, Herzog M, Bojahr A, Leitenberger W, Hertwig A, Shayduk R, and
  Bargheer M, {\em Structural Dynamics} {\bf 1},    (2014).

\bibitem{navi2013a}
Navirian H.~A, Schick D, Gaal P, Leitenberger W, Shayduk R, and Bargheer M,
  {\em Appl. Phys. Lett.} {\bf 104},  021906  (2014).

\bibitem{rein2016}
Reinhardt M, Koc A, Leitenberger W, Gaal P, and Bargheer M, {\em Journal of
  Synchrotron Radiation} {\bf 23},  474  (2016).

\bibitem{schi2013d}
Schick D, Shayduk R, Bojahr A, Herzog M, von Korff~Schmising C, Gaal P, and
  Bargheer M, {\em Journal of Applied Crystallography} {\bf 46},  1372  (2013).

\bibitem{nico2011a}
Nicoul M, Shymanovich U, Tarasevitch A, von~der Linde D, and Sokolowski-Tinten
  K, {\em Appl. Phys. Lett.} {\bf 98},  191902  (2011).

\bibitem{nie2006a}
Nie S, Wang X, Park H, Clinite R, and Cao J, {\em Physical Review Letters} {\bf
  96},  025901  (2006).

\bibitem{jenn1960a}
Jennings L.~D, Miller R.~E, and Spedding F.~H, {\em The Journal of Chemical
  Physics} {\bf 33},  1849  (1960).

\bibitem{gers1969}
Gerstein B.~C, Taylor W.~A, Shickell W.~D, and Spedding F.~H, {\em The Journal
  of Chemical Physics} {\bf 51},  2924  (1969).

\bibitem{vent2014}
Ventura G and Perfetti M, {\em Thermal Properties of Solids at Room and
  Cryogenic Temperatures} (Springer, Netherlands, 2014).

\bibitem{bars1957}
Barson F, Legvold S, and Spedding F, {\em Physical Review} {\bf 105},  418
  (1957).

\bibitem{palm1974}
Palmer S.~B, Lee E.~W, and Islam M.~N, {\em Proceedings of the Royal Society of
  London A: Mathematical, Physical and Engineering Sciences} {\bf 338},  341
  (1974).

\bibitem{nell1969}
Nellis W.~J and Legvold S, {\em Phys. Rev.} {\bf 180},  581  (1969).

\bibitem{koc2017}
Koc A, Reinhardt M, von Reppert A, Leitenberger W, Dumesnil K, Gaal P, Zamponi
  F, and Bargheer M, {\em submitted}  (2017).

\bibitem{Sultan2011}
Sultan M, Melnikov A, and Bovensiepen U, {\em Phys. Status Solidi B} {\bf 248},
   2323  (2011).

\bibitem{doeb2010}
D\"obrich K.~M, Bostwick A, Rotenberg E, and Kaindl G, {\em Phys. Rev. B} {\bf
  81},  012401  (2010).

\end{thebibliography}

\end{document}